\documentclass{aa} % for a printer
\usepackage{graphicx,lscape}
\usepackage{txfonts}
\usepackage{color}
\newcommand{\excs}{\extracolsep{\fill}}
\begin{document}
\title{SSSPM J1102-3431 brown dwarf characterization from accurate proper motion and
trigonometric parallax
\thanks{Based on observations collected at the European Southern
Observatory, Chile (76.C-0543, 077.C-0112, 078.C-0158, 079.C-0229) and at Valinhos meridian circle.}}
\author{R. Teixeira \inst{1,2}
 \and C. Ducourant \inst{2,1}
  \and G. Chauvin \inst{3}
    \and A. Krone-Martins \inst{1,2}
   \and I. Song \inst{4}
    \and B. Zuckerman \inst{5}}

\offprints{teixeira@astro.iag.usp.br}

\institute{
Instituto de Astronomia, Geof\'isica e Ci\^encias Atmosf\'ericas,
       Universidade de S\~ao Paulo,
       Rua do Mat\~ao, 1226 - Cidade Universit\'aria,
       05508-900 S\~ao Paulo - SP,
       Brazil.
          \and
Observatoire Aquitain des Sciences de l'Univers, CNRS-UMR 5804, BP 89, 33270 Floirac, France.
         \and
	  Laboratoire d'Astrophysique, Observatoire de Grenoble,
          414, Rue de la piscine, 38400 Saint-Martin d'H\`eres, France
         \and
	  Spitzer Science Center, IPAC/Caltech, MS 220-6, Pasadena, CA
91125, USA
         \and
	  Department of Physics \& Astronomy and Center for Astrobiology, University of California, Los Angeles, Box 951562, CA 90095, USA}

\date{Received  / Accepted }
\titlerunning {SSSPMJ1102-3431 distance}
%===============================================================================
%           Abstract/Summary
%===============================================================================
\abstract % context heading (optional) % {} leave it empty if
 {In 2005, Scholz and collaborators (Scholz et al. 2005) discovered,
 in a proper motion survey, a young brown dwarf SSSPMJ1102-3431
 (SSSPMJ1102) of spectral type M8.5, probable member of the TW Hydrae
 Association (TWA) and possible companion of the T Tauri star TW Hya.
 The physical characterization of SSSPMJ1102 was based on the
 hypothesis that it forms a binary system with TW Hya.  The recent
 discovery of a probable giant planet inside the TW Hya protoplanetary
 disk with a very short-period (Setiawan et al. 2008) and a disk
 around SSSPMJ1102 (Riaz \& Gizis 2008) make it especially interesting
 and important to measure well the physical parameters of SSSPMJ1102.}
 {Trigonometric parallax and proper motion measurements of SSSPMJ1102
 are necessary to test for TWA membership and, thus, to determine the
 mass and age of this young brown dwarf and the possibility that it
 forms a wide binary system with TW~Hya.}
 {Two years of regular observations at the ESO NTT/SUSI2 telescope,
 have enabled us to determine the trigonometric parallax and proper
 motion of SSSPMJ1102.}
 {With our accurate distance determination of $55.2^{+1.6}_{-1.4}$ pc
 and proper motions of ($-67.2,-14.0$)$\pm0.6$ mas/yr, we could
 confirm SSSPMJ1102 as a very probable member of TWA.  Assuming the TW
 Hydrae association age of $5-10$~Myr, the evolutionary models
 compared to the photometry of this young brown dwarf indicate a mass
 of $\rm{M}=25\pm5~\rm{M}_{\rm{Jup}}$ and an effective temperature
 $T_{\rm{eff}}=2550\pm100$~K.}
 {Our parallax and proper motion determination allow us to precisely
 describe the physical properties of this low mass object and to
 confirm its TWA membership.  Our results are not incompatible with
 the hypothesis that SSSPMJ1102 is a binary companion of  the star TW
 Hya.}
\maketitle
%===============================================================================
%       Corps de article
%===============================================================================

\section{Introduction}
%-------------------------

The TW Hydrae Association (TWA) is a young, nearby association
consisting of about 25 known members. Due to its youth and proximity,
this association has been intensively studied in the last decade
revealing a great variety of systems: tight astrometric binaries good to
calibrate PMS models, stars and brown dwarfs harbouring circumstellar
disks, planetary and brown dwarf companions, and more recently a
putative massive planet embedded in its own proto-planetary
disks (TW Hya; \cite{seti08}). Surprisingly, only five members have
known trigonometric parallaxes. De la Reza et al. (2006)
report a trace-back age of $8.3\pm0.8$~Myr, independent of
evolutionary models. Relying on astrometric and spectroscopic data,
the Galactic space motions of TWA members are traced backward in time
until they occupy a minimum volume in space. This age estimation would
greatly benefit from parallax measurements of additional TWA members.

\cite {scho05} discovered a new young sub-stellar object,
SSSPMJ1102-3431 (SSSPMJ1102), a probable member of TWA. Its
photometric and spectroscopic characteristics suggest a young brown
dwarf of spectral type M8.5. Located $12\arcmin$ from TW Hya and
sharing similar proper motions, Scholz et al. (2005) suggested that
SSSPMJ1102 could form a binary system with TW Hya. Assuming an age of
10 Myr (\cite {webb99}) and the Hipparcos distance for TW Hya they
derived for SSSPMJ1102 a mass of $\approx 25 M_{Jup}$.
Recently, a flat optically thick disk was discovered around
SSSPMJ1102 (\cite {riaz08}) based on a reconstructed mid-infrared
spectral energy distribution using broad-band photometry (\cite {ster04}; 
\cite {riaz06}). Utilizing combined NASA IRTF and Spitzer
spectroscopic observations, \cite {morr08} argued in favor of
high degrees of dust settling to the disk midplane as well as
significant grain growth in the upper layers, suggesting rapid dust
processing compared to disks around stars.

Characterization of SSSPMJ1102 itself and its disk properties
and the question of binarity status with TW Hya make a distance
determination of substantial interest. Since January 2006 we have
conducted astrometric and photometric observations at the ESO NTT
telescope to derive the trignometric parallax of SSSPMJ1102. Our
observations are presented in Section~2. The data reduction and
analysis and the result of this trigonometric parallax programme are
given in Section~3.  Finally, membership in TWA, the physical
properties of SSSPMJ1102 compared to other TWA substellar objects, and
the binarity status with TW Hya are discused respectively in
Sections~4, 5 and 6.

\begin{table}[ht]
\caption{\label{tpi}Astrometric parameters for
SSSPMJ1102-3431 derived in this work. Proper motions and parallax are
absolute quantities.} 
\centerline{
  \begin{tabular}{cccc}
    \hline\hline\noalign{\smallskip}{\smallskip}
$\mu^{*}_{\alpha_{abs}}$(mas/yr)  & $\mu_{\delta_{abs}}$(mas/yr) & $\pi_{abs}$(mas) & $d$(pc)\\
     \hline\noalign{\smallskip}{\smallskip}
-67.2$\pm$0.6&-14.0$\pm$0.6&18.1$\pm$0.5&$55.2^{+1.6}_{-1.4}$\\
     \hline
  \end{tabular}
  }
\end{table}

\begin{table}[ht]
\caption{\label{phot}Bessel (V, R and I derived in this work) and 2MASS (J, H and K, \cite{cutr03}) 
apparent and absolute magnitudes for SSSPMJ1102-3431.} 
\centerline{
  \begin{tabular}{rrr}
    \hline\hline\noalign{\smallskip}{\smallskip}
 &\multicolumn{1}{c}{m(mag)} & \multicolumn{1}{c}{M(mag)}\\
      \hline\noalign{\smallskip}{\smallskip}
V & 21.46 $\pm$ 0.04 & 17.75 $\pm$ 0.19\\ 
R & 19.14 $\pm$ 0.04 & 15.43 $\pm$ 0.19\\
I & 17.90 $\pm$ 0.03 & 14.19 $\pm$ 0.18\\
J & 13.034 $\pm$ 0.024 & 9.32 $\pm$ 0.17\\
H & 12.356 $\pm$ 0.022 & 8.65 $\pm$ 0.17\\
K & 11.887 $\pm$ 0.024 & 8.18 $\pm$ 0.17\\
     \noalign{\smallskip}\hline
  \end{tabular}
  }
\end{table}

\section{Observations}
%--------------------------

Astrometric and photometric (V, R, I) observations were performed
between 2006 and 2007 in direct imaging mode with the ESO NTT-SUSI2
instrument that ensures a good compromise between a large field of
view $(5.5\arcmin x 5.5\arcmin)$ for a sufficient sample of background
stars and a small pixel size ($80.5$ mas) necessary to reach sub-milli
arc second astrometric precision.

Data at six observational epochs were acquired with a total of twelve
nights of observation. All observations were done around transit
to minimize the differential color refraction effects (DCR). Multiple
exposures were obtained at each epoch to reduce astrometric errors
and enhance S/N.  An I-band filter was selected to minimize the DCR
effect. Residual DCR effects were removed from single observations
following methods described in \cite {duco07}.

The alignment of CCD axes and the scale determination were obtained
using the 2MASS catalogue (\cite {cutr03}) and from CCD meridian
observations at Valinhos Observatory (\cite {teix00}).

All image frames were measured using the {\tt DAOPHOT II} package
(\cite{stet87}), fitting a stellar point-spread function for each
frame. Finally we created catalogs of measured positions $(x,y)$,
internal magnitudes, and associated errors for all stars on each
frame.

\section{Trigonometric parallax determination}
%--------------------------------------------------------

These observational data were processed through a global treatment as
described in \cite {duco07} and \cite {duco08} and a solution was
derived for SSSPMJ1102, relative to background stars (I$\le22.5$ mag)
: $\pi_{rel}=17.5\pm0.6$ mas, $\mu^{*}_{\alpha_{rel}}=-61.0\pm0.6 $
mas/yr, $\mu_{\delta_{rel}}=-15.0\pm0.6$ mas/yr. Then a statistical
conversion from relative to absolute parallax and proper motions,
based on the Besan\c con Galaxy model (\cite{robi03}, \cite{robi04}),
was derived ($\Delta\pi=0.6$ mas, $\Delta\mu^{*}_{\alpha}=-6.2$ mas/yr,
$\Delta\mu_{\delta}=+1.0$ mas/yr). We give in Table \ref{tpi} absolute
astrometric parameters and in Table \ref{phot} the photometry for SSSPMJ1102 obtained in
this work.

\begin{figure}[ht]
  \includegraphics*[angle=0,width=8cm]{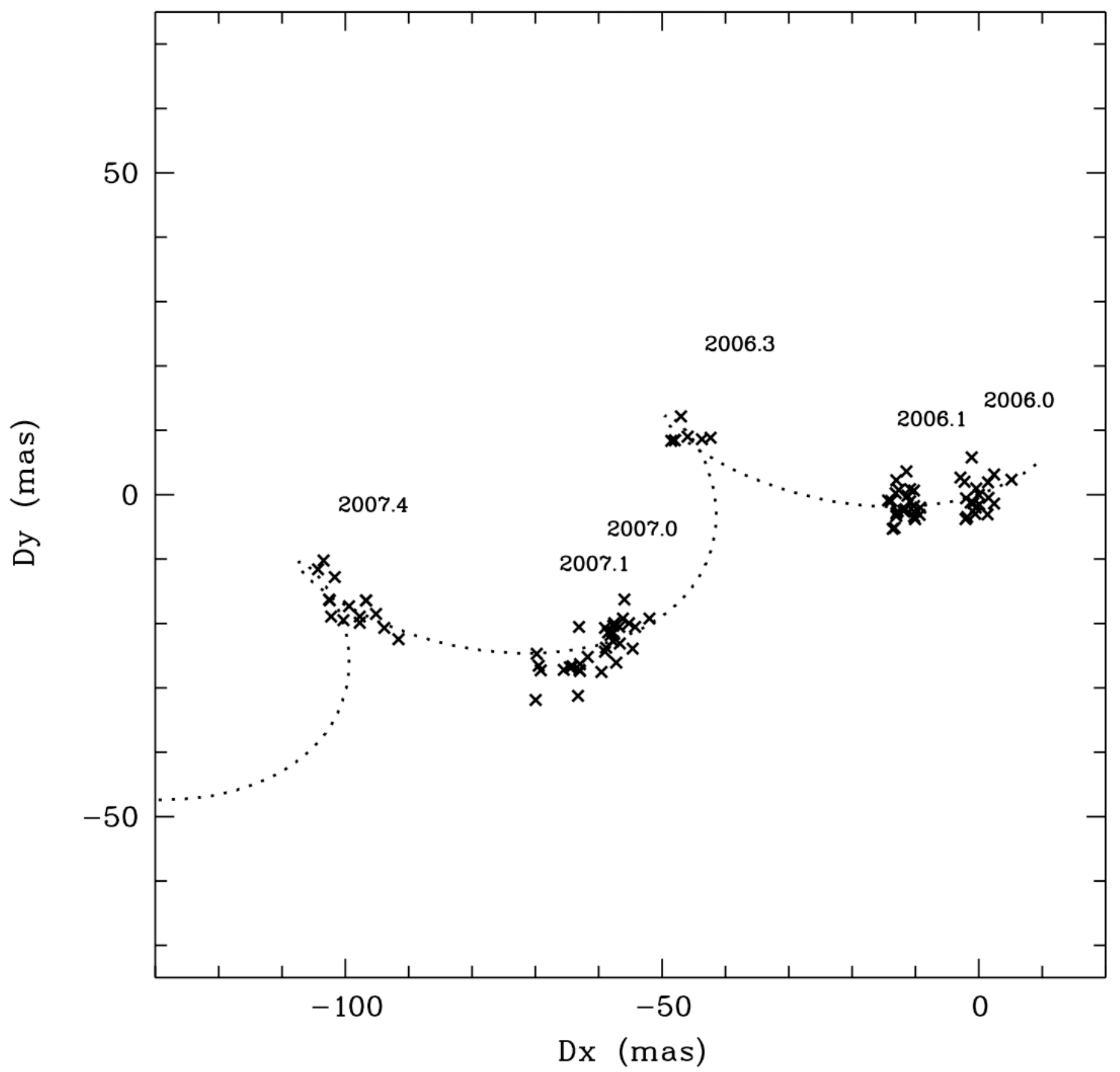}
  \caption{\label{pi}Observations together with the fitted path
due to parallax and proper motion for SSSPMJ1102. The dashed curve
represents the least squares solution.}
\end{figure}

We present in Figure \ref{pi} the observations of SSSPMJ1102, together
with the fitted path (relative parallax and proper motions).

%FIGURE Trig parallax

\section{Membership in TWA}
%----------------------------------------------

The distance derived in this work (d=$55.2^{+1.6}_{-1.4}$ pc) is well
within the limits of the boundaries of the photometrically derived
distances of the 25 TWA members (\cite{zuck04})~: 20~pc to 130~pc and
also in good accordance with the boundaries of the accurate HIPPARCOS
parallaxes (\cite {esa97}) of 5 members : 47 to 104 pc.

The proper motions obtained in this work agree, at about the 1$\sigma$
level, with those obtained by \cite{scho05} ($-82\pm12$ mas/yr,$-12\pm
6$ mas/yr) and are fully compatible with the proper motions of TWA
members ($\mu_\alpha^{*}$ from -33 to -122 mas/yr, $\mu_\delta$ from
-10 to -43 mas/yr).

Following the \cite{debr99} convergent point method we can test the
membership of SSSPMJ1102 in TWA.  Using our proper motion results, and
the estimate of TWA's convergent point obtained by \cite{mama05}
($\alpha = 103^{o}.2 \pm 1^{o}.5$, $\delta = -30^{o}.7 \pm 1^{o}.5$),
we find that most of SSSPMJ1102 proper motion's magnitude is
concentrated in the direction toward the convergent point ($\mu_{\|} =
-68.55 \pm 0.61$ mas yr$^{-1}$) and that its perpendicular component
is small ($\mu_{\bot} = 3.06 \pm 2.32$ mas yr$^{-1}$), being
compatible with zero, as expected for an ideal cluster member.

The accordance of both our distance estimate and our proper motions
with the characteristics of TWA strongly supports the membership of SSSPMJ1102
in the TWA Association.

\section{Physical properties}
%----------------------------------------------

Considering SSSPMJ1102 as a member of TWA, one can use evolutionary
model predictions to derive the physical properties of this young M8.5
brown dwarf. We compared its 2MASS JHK photometry (converted into the
CIT system) with the predictions of the DUSTY evolutionary
models. The predicted mass, effective temperature, surface gravity and
luminosity are reported in Table~\ref{tab:models}.  They are compared
to the (model-dependent) physical properties of the five other
substellar dwarf objects of TWA, covering spectral types between M8 to
mid-L and part of a large variety of systems: isolated (2M1139, \cite {gizi02}; 
DENIS\,1245, \cite {loop07}), in a binary (2M1207\,AB,
\cite {chauvin2004}, \cite {chauvin2005}) or as a companion to a stellar primary
(TWA5\,B, \cite {lowr99}). Uncertainties in the mass of
2M1207\,B are discussed by \cite {mohanty2007}, \cite {gizis2007}, \cite {mamajek2007}
and \cite {duco07}. Distances
(sometimes proper motion) remain uncertain for TWA5, 2M1139 and
DENIS\,1245 and would benefit from future parallax measurements.

\begin{table}[t]
\caption{\label{tab:models}Physical properties of SSSPMJ1102,
predicted by the DUSTY evolutionary models for a given age of
$5-10$~Myr. They are compared to the properties of the five other
substellar systems in TWA using the same model, the same age and
trigonometric or photometric (in parenthesis) parallaxes. In the case
of 2M1207\,B, the COND model predictions are also considered as
discussed in \cite {duco07}.}
\begin{tabular*}{\columnwidth}{@{\excs}lllllll}
\hline\hline\noalign{\smallskip}
Name           &      d              &  SpT     &  Mass                 &   $T_{\rm{eff}}$  &  $\rm{log(L/L}_{\odot})$    \\
              &    (pc)             &          & ($\rm{M}_{\rm{Jup}}$) & (K)               &                             \\
\noalign{\smallskip}\hline\noalign{\smallskip}
2M1207\,A      &  $52.4$             &  M8      & $25\pm5$              &    $2550\pm150$   &       $-2.7\pm0.2$          \\
2M1139         &  (46)               &  M8      & $25\pm5$              &    $2550\pm150$   &       $-2.6\pm0.3$          \\
TWA5\,B        &  (45)               &  M8.5    & $25\pm5$              &    $2550\pm150$   &       $-2.6\pm0.3$          \\
\noalign{\smallskip}
SSSPMJ1102     &  $55.2$             &  M8.5    & $25\pm5$              &    $2550\pm150$   &       $-2.6\pm0.3$          \\
\noalign{\smallskip}
DENIS\,1245    &  (90)               &  M9.5    & $15-20$               &    $2350\pm150$     &     $-2.9\pm0.3$          \\
2M1207\,B      &   $52.4$            &  late-L  & $8\pm2$               &    $1600\pm100$   &       $-3.8\pm0.1$          \\
               &  $52.4$             &  late-L  & $4\pm1$               &    $1150\pm150$   &       $-4.5\pm0.2$          \\

\noalign{\smallskip}\hline
\end{tabular*}
\end{table}

\section{Binarity}
%----------------------------------------------
\cite{scho05} suggested that SSSPMJ1102 forms a wide binary system
with TW Hya, but with the lack of distance and accurate proper motion
for SSSPMJ1102, the authors could not draw any firm conclusions.

The distance and proper motions derived here for SSSPMJ1102 (d=55.2$^{+1.6}_{-1.4}$
pc, (-67.2, -14.0) $\pm$ 0.6 mas/yr) can be compared with the recent "Hipparcos
new reduction" catalogue (\cite {vanl07a}, \cite {vanl07b}) data for TW Hya
(d=55.4$^{+11.4}_{-8.1}$ pc, -68.2 $\pm$ 2.7 mas/yr, -13.2 $\pm$ 2.2 mas/yr). We observe that both
objects are at the same distance from the sun and have very similar
proper motions.

Considering SSSPMJ1102 as a member of TWA, from our astrometric
measurements and the convergent point from \cite {mama05}, we can
estimate the SSSPMJ1102 radial velocity $V_{r}=14.13\pm 0.79$ km/s
leading to the spacial velocity components: (-11.4, -19.3, -4.7) km/s, very
similar to that of TW Hya's (-11.7, -17.9, -4.9) km/s.

From the coordinates determined here for SSSPMJ1102 
\textbf{ ($11^{h} 02^{m} 09.803^{s}$, $-34 \degr30\arcmin35.56\arcsec$, 2006.653) }and the
position and proper motions given in the new reduced Hipparcos for TW
Hya ($11^h 01^m 51.954^s, -34\degr 42\arcmin 16.91\arcsec$, 1991.25),
we derive an angular separation in 2006.653 of 12.3 $\arcmin$. From
this angular separation and the \textbf{ distance of SSSPMJ1102 (55.2$^{+1.6}_{-1.4}$ pc, this
work) we derive a projected separation on the sky of about 41000AU.}

The good agreement between the kinematic characteristics of both
objects show that in addition to occupying similar positions in space,
they clearly share the same spatial movement. However the large
separation does not support the hypothesis that these two objects are
gravitationally bound. Complementary observations are necessary to
clarify this point.

\section{Conclusions}
%----------------------

Motivated by the need to have accurate distance determinations for
members of the TW Hydrae Association and by specific interest in the
brown dwarf SSSPMJ1102, we measured its trigonometric parallax with an
error $<$3$\%$, thus ensuring a precise distance determination and a
refined physical characterization of this sub-stellar object. The good
accordance of our precise distance and proper motions of SSSPMJ1102
with those characterizing TWA ensure that this object belongs to the
TWA Association.

Our results show that SSSPJ1102 and TW Hya are in the same region of
the space sharing the same spatial movement as suspected by \cite {scho05}
but this fact still does not allow us to conclude if these
two objects belong to a wide binary system or not.

\begin{acknowledgements} 
We would like to thank the staff of ESO-VLT and CFHT and Gilles
Chabrier, Isabelle Baraffe and France Allard for providing the latest
update of their evolutionary models. We also acknowledge partial
financial support from the {\sl Programmes Nationaux de Plan\'etologie
et de Physique Stellaire} (PNP \& PNPS) (in France),  the Brazilian
Organism FAPESP and CAPES and French Organism COFECUB.
\end{acknowledgements}

\end{document}